\documentclass[11pt,twoside]{article}

\usepackage{asp2006}
\usepackage{epsf}
\usepackage{lscape}
\usepackage{graphicx}
\begin{document}
\title{Polarization of Quasars: Rotated and Funnel-shaped Outflow}
\author{Hui-Yuan Wang, Ting-Gui Wang and Jun-Xian Wang}
\affil{Center for Astrophysics,
University of Science and Technology of China,
Hefei, 230026, China}    

\begin{abstract} 
Polarization is a useful probe to investigate the geometries and
dynamics of outflow in BAL QSOs. We perform a Monte-Carlo
simulation to calculate the polarization produced by resonant and
electron scattering in BALR. We find: 1)A rotated and
funnel-shaped outflow is preferred to explain many observed
polarization features. 2)The resonant scattering can contribute a
significant part of NV emission line in some QSOs.
\end{abstract}



\section{Introduction}
Broad absorption lines (BALs) are observed in the UV spectra of
about 10-20\% QSOs and taken as an important indicator for the
central engine of QSOs. BAL carries only information of the BAL
region along the line of sight to the continuum source while the
global information is contained in the scattered polarized light.
BAL QSOs are the only highly polarized population among radio
quiet QSOs. Ogle et al. (1999) studied the polarization of 36 BAL
QSOs and found: position angle(PA) rotation in the BALs,
blueshifted BALs in polarized flux relative to that in total flux
spectra and deficit of polarized flux redward of the BALs. In
order to explain these features we perform a Monte-Carlo
simulation to calculate resonant and electron scattering for
radial outflows of two different geometries: equatorial and
funnel-shaped outflows with and without rotational velocity. We
refer the readers to Wang et al.(2005, 2007) for details.

\section{Results}

We present our results in Fig 1(left) for rotated outflow with
geometry similar to Elvis(2000, his Figure 3). In an axisymmetric
model, PA rotation in the BALs can be produced only when the
outflow carries angular momentum. The PA rotation increases with
the angular velocity as well as the subtending angle of the flow.
In order to produce a PA rotation $>$ 10\deg observed in a few BAL
QSOs, subtending angle of the outflow should be larger than
25\deg. Similar requirement is imposed to explain the red deficit
of polarized flux observed in some BAL QSOs. If the subtending
angle of outflow is larger than 25\deg the polarization plane of
back-scattered light is perpendicular to that of continuum and the
polarization vectors of the light redward of the BAL cancel,
giving a deficit of polarized flux(Wang et al. 2007). Assuming the
electron scattering region(for example the shielding gas, Murray
et al. 1995, Proga et al. 2000) is cospatial with the outflow, we
calculate the effect of electron scattering to the polarization in
BALs. Because of rotation of the outflow and similar size of two
scattering regions, the electron scattered light fill the trough
at $v> -v_\varphi$, where $v_\varphi$ is the rotational velocity
of outflow at the launching radius, and the BAL in polarized flux
is blueshifted relative to that in the total flux, just as shown
in Fig. 1(right).

We also find resonantly scattered light will contribute a
significant part of NV line that is widely used to determine the
metal abundance of QSOs. For an equatorial outflow the maximum
ratio between the equivalent width(EW) of the scattering emission
to the BAL EW increases from 11\% to 18.6\% if the covering factor
of BALR increases from 20\% to 40\%. The scattering emission of NV
is even stronger because of the scattering of the strong
Ly$\alpha$ emission line. Therefore, a strong NV may not indicate
a high metal abundance but a large covering factor of the
BALR(Wang et al. 2007).

\begin{figure}
\includegraphics[scale=0.95]{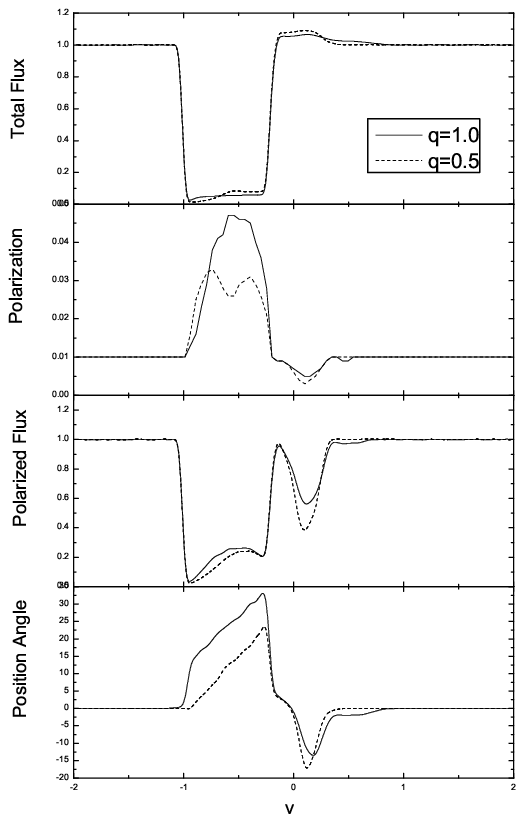}\includegraphics[scale=0.95]{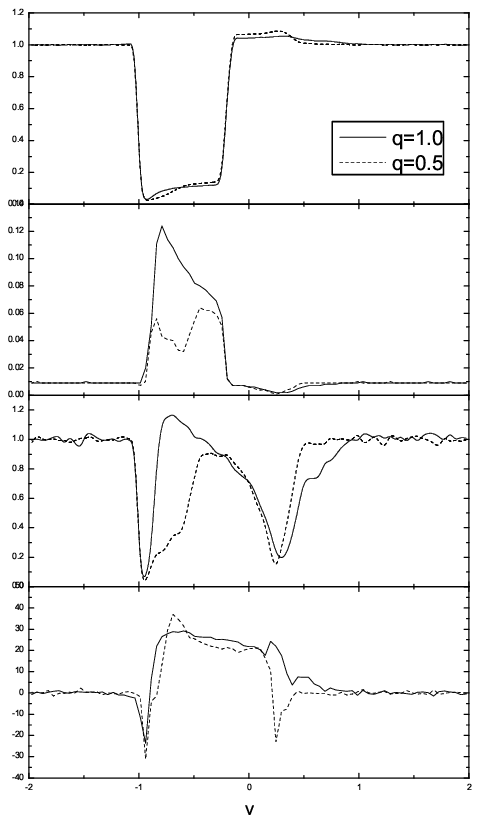}
\caption{The total flux, polarization degree, polarized flux and
position angle. Here $q=v_\varphi/v_0$, where $v_0$ is the radial
terminal velocity of outflow. In the left panel we only calculate
the resonant scattering and in the right panel we consider both
resonant and electron scattering.}
\end{figure}




\end{document}